\documentclass[11pt]{amsart}

\usepackage{amsmath}
\usepackage{amsfonts}
\usepackage{amssymb}
\usepackage{amscd}
\usepackage{amsthm}
\usepackage{framed}
\usepackage{fullpage}
\usepackage{graphicx}
\usepackage{latexsym}
\usepackage[numbers]{natbib}  

\usepackage{hyperref}
\hypersetup{colorlinks,linkcolor={red},citecolor={blue},urlcolor={blue}}

\newtheorem{theorem}{Theorem}[section]

\newtheorem{conjecture}[theorem]{Conjecture}

\theoremstyle{definition}

\numberwithin{equation}{section}

\newcommand{\HH}{\mathbb H}

\newcommand{\NN}{\mathbb N}

\newcommand{\QQ}{\mathbb Q}

\newcommand{\ZZ}{\mathbb Z}

\newcommand{\SL}{\mathop{\mathrm {SL}}\nolimits}

\newenvironment{psmallmatrix}
  {\left(\begin{smallmatrix}}
{\end{smallmatrix}\right)}

\begin{document}

\title[Dynkin diagrams, generalized Nahm sums and 2d CFTs]{Dynkin diagrams, generalized Nahm sums and 2d CFTs}

\author{Kaiwen Sun}
\address{School of Mathematical Sciences, University of Science and Technology of China, Hefei 230026, Anhui, China}
\email{kwsun@ustc.edu.cn}

\author{Haowu Wang}
\address{School of Mathematics and Statistics, Wuhan University, Wuhan 430072, Hubei, China}
\email{haowu.wangmath@whu.edu.cn}


\date{\today}


\begin{abstract} 
A folklore conjecture states that the Nahm sum associated with a pair of Dynkin diagrams of type $ADET$ is a modular function. In this paper, we extend this conjecture to Dynkin diagrams of type $ABCDEFGT$ in the context of generalized Nahm sums. The modular Nahm sums are closely related to the characters of 2d rational conformal field theories. In this work, we identify many specific generalized Nahm sums with characters of some well-studied 2d CFTs. For example, we find that the generalized Nahm sums associated with $(T_1, C_r)$ and $(T_1,D_r)$ correspond to the supersymmetric Virasoro minimal models $\mathrm{SM}(4r+6, 4)$ and $\mathrm{SM}(8r+4, 2)$, respectively. 
\end{abstract}

\maketitle

\section{Introduction}
We begin with a few standard notation. Let $\NN$ denote the set of non-negative integers. We set
$$
(x;q)_0=1 \quad \text{and} \quad  (x;q)_n=\prod_{j=0}^{n-1}(1-xq^j)
$$
for any positive integer $n$, and define
\begin{align*}
(x;q)_\infty&=\prod_{j=0}^{\infty}(1-xq^j),\\   
(x_1,\cdots,x_m; q)_n &= (x_1;q)_n\cdots (x_m;q)_n, \quad n\in \NN \cup \{\infty\}.
\end{align*}
The famous Rogers--Ramanujan identities state that
$$
G(q)=\sum_{n=0}^\infty \frac{q^{n^2}}{(q;q)_n}=\frac{1}{(q,q^4;q^5)_\infty}, \qquad H(q)=\sum_{n=0}^\infty \frac{q^{n^2+n}}{(q;q)_n}=\frac{1}{(q^2,q^3;q^5)_\infty}.
$$
Remarkably, $q^{-\frac{1}{60}}G(q)$ and $q^{\frac{11}{60}}H(q)$ turn out to be exactly the two characters of the Lee--Yang model, namely the Virasoro minimal model $\mathrm{M}(5,2)$, one of the simplest  rational two-dimensional conformal field theories (2d CFTs). In particular, they form a vector-valued modular function for a certain two-dimensional complex representation of $\SL_2(\ZZ)$ and are therefore modular functions. 

In the 1990s, physicists \cite{NRT93, KM93, KKMM93a, KKMM93b, Mel94, BG95} generalized the above infinite sums in the study of rational 2d CFTs. Fix a positive integer $r$, a positive-definite symmetric matrix $A\in\QQ^{r\times r}$, a vector $B\in \mathbb{Q}^r$, and a scalar $C\in\mathbb{Q}$. The $r$-fold $q$-hypergeometric series
\begin{equation}
f_{A,B,C}(\tau):=\sum_{n=(n_1,...,n_r)\in\NN^r} \frac{q^{\frac{1}{2}n^tAn+n^tB+C}}{(q;q)_{n_1}\cdots (q;q)_{n_r}}, \quad \tau\in \HH, \; q=e^{2\pi i\tau}
\end{equation}
is termed a \emph{Nahm sum} associated with the triple $(A,B,C)$. It was observed that the fermionic sum representations for characters of some well-known 2d CFTs induce Nahm sums that are modular forms. Analyzing the asymptotic expansion for $\tau$ tending to $i\infty$, the weight of $f_{A,B,C}$ has to be zero if it is a modular form for a congruence subgroup of $\SL_2(\ZZ)$ (see, e.g., \cite[Lemma 3.1]{VZ11}). Therefore, the modular Nahm sums are modular functions with non-negative integral Fourier coefficients, so they are likely to connect to the characters of rational 2d CFTs. This illustrates why we are particularly interested in modular Nahm sums. We call $(A,B,C)$ a \emph{modular triple} if $f_{A,B,C}$ is modular. Nahm sums also have wide applications to partitions \cite{Gor61, And74}, knots \cite{AD11, GZ24, Sto24,Kucharski:2017ogk}, Chern--Simons theory \cite{KS24}, and 3d topological field theories \cite{Gang:2023rei,GKPS24}.  

Nahm \cite{Nah07} proposed a conjectural criterion for a matrix $A$ such that $f_{A,B,C}$ is modular for some pair $(B,C)$. Let $a_{ij}$ be the entries of $A$. The algebraic equation 
\begin{equation*}
1-X_i = \prod_{j=1}^r X_j^{a_{ij}}, \quad i=1,2,...,r    
\end{equation*}
is called the \emph{Nahm equation}. Each solution $X$ to the Nahm equation yields an element $[X]$ in the Bloch group $B(\mathbb{C})$. Note that the Nahm equation has a unique solution in $(0,1)^r$ (see, e.g., \cite[Lemma 2.1]{VZ11}) and we denote the corresponding element in $B(\mathbb{C})$ by $\xi_A$. We fix $A$ as above and consider the following three conditions:
\begin{itemize}
\item[(a)] the class $[X]\in B(\mathbb{C})$ vanishes for all solutions $X$ of the Nahm equation;
\item[(b)] the special class $\xi_A\in B(\mathbb{C})$ vanishes;
\item[(c)] there exist $B\in\QQ^r$ and $C\in\QQ$ such that $f_{A,B,C}$ is modular. 
\end{itemize}

Zagier \cite{Zag07} found that (b) does not imply (a) or (c) in general. Vlasenko--Zwegers \cite{VZ11} showed that (c) does not imply (a) by counterexamples. Calegari--Garoufalidis--Zagier \cite{CGZ23} proved that (c) implies (b). It is believed that (a) implies (c). This prediction is sometimes referred to as Nahm's conjecture, although it was stated by Zagier \cite{Zag07} as a potential precise formulation of Nahm's insight. Remark that Nahm's conjecture remains widely open, and it was only proven in the rank-one case by Zagier \cite{Zag07}. 

It has long been known and conjectured that Cartan matrices of type $ADET$ produce modular Nahm sums. This folklore conjecture in theoretical physics was inspired by the study of fermionic sum representations for characters of coset 2d CFTs (see \cite{KM93, KKMM93a, KKMM93b, Mel94, BG95, KN11}). Let $X$ and $Y$ be Dynkin diagrams of type $ADET$. We denote by $C(X)$ and $C(Y)$ the corresponding Cartan matrices, respectively. Recall that $T_r$ is the Tadpole diagram $A_{2r}/\ZZ_2$ introduced in \cite{KM90}, which is related to the twisted affine Lie algebra $A_{2r}^{(2)}$. The Cartan matrix $C(T_r)$ is identical to that of $A_r$ except for the entry $C(T_r)_{rr}=1$, and the Coxeter number of $T_r$ is $h(T_r)=2r+1$. We define 
\begin{equation}
A(X,Y)=C(X)\otimes C(Y)^{-1},    
\end{equation}
where $\otimes$ is the Kronecker product of two matrices. Recall that the Kronecker product of the square matrices $A=(a_{ij})_{i,j=1}^n$ and $B=(b_{ij})_{i,j=1}^m$ is a square matrix of order $mn$ defined as 
$$
A\otimes B = \begin{pmatrix}
a_{11}B & \cdots & a_{1n}B \\
\vdots & \ddots & \vdots \\
a_{n1}B & \cdots & a_{nn}B 
\end{pmatrix}.
$$
The folklore conjecture states that the Nahm sum associated with $\big(A(X,Y),0,C(X,Y)\big)$ is modular, here $C(X,Y)=-c(X,Y)/24$ and 
\begin{equation}
c(X,Y)=\frac{r(X)r(Y)h(X)}{h(X)+h(Y)}    
\end{equation}
indicates the central charge of the corresponding coset 2d CFT, where $r(-)$ and $h(-)$ denote the rank and the Coxeter number of a Dynkin diagram, respectively. In \cite{Lee13} it was proved that every solution of the Nahm equation associated with $A(X,Y)$ induces an element that vanishes in $B(\mathbb{C})$. Therefore, Nahm's conjecture would yield this folklore conjecture. 

How to generalize this conjecture to other types of Dynkin diagram is a natural question. In this paper, we address this problem in the context of generalized Nahm sums.  
Recently, Mizuno \cite{Miz23} introduced Nahm sums for symmetrizable matrices and extended Nahm's conjecture to this case. Let $D=\mathrm{diag}(d_1,...,d_r)$ be a diagonal matrix and $A\in \mathbb{Q}^{r\times r}$ such that $AD$ is symmetric and positive-definite, where the diagonal entries $d_j$ are all positive integers. Let $B\in \QQ^r$ and $C\in \QQ$. The generalized Nahm sum associated with the quadruple $(A,B,C,D)$ is defined as
\begin{equation}
f_{A,B,C,D}(\tau):=\sum_{n=(n_1,...,n_r)\in\NN^r} \frac{q^{\frac{1}{2}n^tADn+n^tB+C}}{(q^{d_1};q^{d_1})_{n_1}\cdots (q^{d_r};q^{d_r})_{n_r}}.    
\end{equation}
Similarly, by the asymptotic formula (see \cite[Theorem 2.1]{Miz23}), if $f_{A,B,C,D}(\tau)$ is a modular form for a congruence subgroup of $\SL_2(\ZZ)$, then its weight must be $0$. The quadruple $(A,B,C,D)$ is called modular if the associated generalized Nahm sum is modular. Our conjecture is as follows. 
\begin{conjecture}\label{Conj:main}
Let $X$ and $Y$ be Dynkin diagrams of type $ABCDEFGT$ with Cartan matrices $C(X)$ and $C(Y)$, respectively. Let $D(X)$ denote the diagonal matrix whose $j$-th diagonal entry is defined as the quotient $\alpha_j^2/\beta^2$ of the square lengths of the simple roots, where $\alpha_j$ is the $j$-th simple root and $\beta$ is a short root.  We set 
\begin{equation}
A(X,Y)=C(X)\otimes C(Y)^{-1} \quad \text{and} \quad D(X,Y)=D(X)\otimes D(Y),
\end{equation}
where $\otimes$ is the Kronecker product of two matrices. We also set 
$$
C(X,Y)=-c(X,Y)/24, 
$$
where the central charge $c(X,Y)$ is given by
\begin{equation}\label{eq:cXY}
c(X,Y)=\frac{\mathrm{tr}\big(D(X,Y)\big)h(X)}{h(X)+h(Y)},   
\end{equation}     
and $h(-)$ is the Coxeter number. Then $\big(A(X,Y),0,C(X,Y),D(X,Y)\big)$ is a modular quadruple. 
\end{conjecture}

The generalized Nahm sum associated with the above quadruple is well-defined because
$$
A(X,Y) D(X,Y)=\big(C(X)D(X)\big)\otimes \big(C(Y)^{-1}D(Y)\big)
$$
is symmetric, as $C(X)D(X)$ is symmetric and $C(Y)^{-1}D(Y)=D(Y)\big(D(Y)^{-1}C(Y)^{-1}\big)D(Y)$. 

Zagier \cite{Zag07} proposed a conjectural duality between modular triples. Mizuno \cite[Conjecture 4.1]{Miz23} generalized Zagier's conjecture to modular quadruples, which states that if the quadruple $(A,B,C,D)$ is modular, then the following dual quadruple should also be modular
\begin{equation}
(A^*,\, B^*,\, C^*,\, D^*):=\Big( A^{-1},\, A^{-1}B,\, \frac{1}{2}B^t(AD)^{-1}B-\frac{\mathrm{tr}(D)}{24} - C,\, D \Big).
\end{equation}
Recently, Wang \cite{Wan24} discovered counterexamples to Zagier's duality conjecture and its generalization. Fortunately, Conjecture \ref{Conj:main} shows that Zagier's duality may still hold when $B=0$. Let $A$ and $B$ be square matrices. It is known that $(A\otimes B)^{-1}=A^{-1}\otimes B^{-1}$ and $A\otimes B=P^t(B\otimes A)P$ for some permutation matrix $P$. Therefore, the dual of $\big(A(X,Y),0,C(X,Y),D(X,Y)\big)$ corresponds to $\big(A(Y,X),0,C(Y,X),D(Y,X)\big)$ up to permutation. 

By \cite[Proposition 3.5.1]{Sto24}, generalized Nahm sums can be reduced to ordinary Nahm sums by increasing the rank. Therefore, Conjecture \ref{Conj:main} also generates infinitely many new candidates of modular ordinary Nahm sums. 
If one could extend the result of \cite{Lee13} to all Dynkin diagrams, then the generalization of Nahm's conjecture would yield Conjecture \ref{Conj:main}. 

Conjecture \ref{Conj:main} is known to hold in many cases. For example, the famous Andrews--Gordon identity \cite{Gor61,And74} and the Bressoud identity \cite{Bre79} show that the generalized Nahm sums associated with pairs $(A_1, T_r)$ and $(A_1, C_r)$ are modular, respectively. In addition, Conjecture \ref{Conj:main} generates $35$ generalized Nahm sums of rank less than $4$ in total, and $28$ of them have been shown to be modular. We will discuss in detail what is known about this conjecture in Section \ref{sec:known-cases}.  

In the 1990s, it was noticed that the Nahm sums associated with pairs of type $(A_1,Y)$ with $Y$ as simply-laced Dynkin diagram are related to the characters of the coset 2d CFT $\big({(Y)_1\times (Y)_1}\big)/{(Y)_2}$ (see, e.g., \cite{KKMM93a,KKMM93b}). More generally, it is possible that all modular Nahm sums come from characters of rational 2d CFTs. For example, the Nahm sums associated with $(A_1,T_r)$ give the characters of the Virasoro minimal model $\mathrm{M}(2r+3,2)$, and the Nahm sums associated with $(T_1,T_{2r})$ correspond to the characters of the supersymmetric Virasoro minimal model $\mathrm{SM}(4r+4,2)$. Some isolated such correspondences can be found in \cite{GKPS24}. In this paper, we further identify many specific generalized Nahm sums associated to pairs of Dynkin diagrams with some well-studied 2d CFTs. More precisely, we express the Nahms sums as $\ZZ$-linear combinations of characters of some rational 2d CFTs. We summarize the correspondences in Table \ref{tb:full} below.

\begin{table}[ht]
\def\arraystretch{1.2}
\[
\begin{array}{l|c|c|c}
(X,Y) & c & \mathrm{CFT} & \mathrm{ref} \\\hline 
(A_1,A_r) &  2r/(r+3) &  \ZZ_{r+1}\ \textrm{parafermion} &  \cite{KKMM93a} \\
(A_1,B_r) & (2r-1)/(r+1)   &  ? &  \cite{BKRS23}  \\
(A_1,C_r) &  1  &  U(1)_{4(r+1)} &   \eqref{eq:A1Cr} \\
(A_1,D_r) & 1  & U(1)_{r}  &  \cite{KKMM93a} \\
(A_1,E_6) &  6/7 & \mathrm{M}_{\rm sub}(7,6)  & \cite{KKMM93a}  \\
(A_1,E_7) &  7/10 & \mathrm{M}(5,4)  & \cite{KKMM93a}  \\
(A_1,E_8) & 1/2  &  \mathrm{M}(4,3) & \cite{KKMM93a, WP94}  \\ 
(A_1,F_4) &  6/7  & \mathrm{M}(7,6)  &  \eqref{eq:A1F4}\eqref{eq:A1F4b}  \\
(A_1,G_2) &  1  & U(1)_{36}  &  \eqref{eq:A1G2}  \\
(A_1,T_r) &  2r/(2r+3)  & \mathrm{M}_{\rm eff}(2r+3,2)  &  \cite{Gor61,And74}    \\
\hline
(T_1,A_r) &  3r/(r+4)  & ?  &  - \\
(T_1,B_r) &  3(2r-1)/(2r+3)  &  ? &  -  \\
(T_1,C_r) &  {3(r+1)}/{(2r+3)}  & \mathrm{SM}_{\rm eff}(4r+6,4)  &  \eqref{eq:T1Cr}  \\
(T_1,D_r) & 3r/(2r+1)  &  \mathrm{SM}_{\rm eff}(8r+4,2) &  \eqref{eq:T1Dr}  \\
(T_1,E_6) &  6/5 &  \mathrm{M}_{\rm eff}(5,2)\otimes \mathrm{M}_{\rm sub}(6,5)  &  \eqref{eq:T1E6}  \\
(T_1,E_7) &  1 &  ?  &  - \\
(T_1,E_8) &  8/11 &  \mathrm{M}_{\rm eff}(11,2) &  \eqref{eq:T1E8} \\ 
(T_1,F_4) &  6/5  & \mathrm{M}_{\rm eff}(5,2)\otimes \mathrm{M}(6,5)  &  \eqref{eq:T1F4}\eqref{eq:T1F4b}  \\
(T_1,G_2) &  4/3  &  ? &  \cite{KR15}  \\
(T_1,T_{2r}) &   3r/(2r+2) &  \mathrm{SM}_{\rm eff}(4r+4,2) &  \cite{Mel94} \\
(T_1,T_{2r+1}) &  3(2r+1)/(4r+6)  & F\otimes \mathrm{M}_{\rm eff}(2r+3,2)  &  \cite{War03} \\
\hline
(A_{2r-1},T_k) & 2kr(2r-1)/(2(k+r)+1) & \mathfrak{osp}(1|2r)_k &   \cite{CG25} \\
(E_8,T_1) & 80/11  &  (E_8)_1/\mathrm{M}_{\rm eff}(11,2) & \eqref{eq:E8T1} \\
(T_1,A_1) & 3/5 & \mathrm{M}_{\rm eff}(5,3) & \eqref{eq:T1A11}\eqref{eq:T1A12} \\ 
(T_1,A_2) & 1 & U(1)_{3}  & \eqref{eq:T1A2} \\
(T_1,A_3) & 9/7 & \mathrm{SM}_{\rm eff}(28,2) & \eqref{eq:T1A3} \\
(T_2,T_1) &  5/4 &  \mathrm{SM}_{\rm eff}(8,2)\otimes F &  \eqref{eq:T2T11}\text{-}\eqref{eq:T2T14} \\
(T_2,A_1) & 10/7 & \mathrm{SM}_{\rm eff}(84,2) & \eqref{eq:T2A1-1}\text{-}\eqref{eq:T2A1-3} \\
(A_2,T_1) & 1 & (A_1)_1 & \eqref{eq:A2T1} \\ 
(A_1,A_2) & 4/5 & \mathrm{M}_{\rm sub}(6,5) & \eqref{eq:A1A21}\eqref{eq:A1A22} \\
(A_2,A_1) & 6/5 & D_{\rm 2A} & \eqref{eq:A2A11}\eqref{eq:A2A12} \\
(A_1,A_3) & 1 & U(1)_{3}  & \eqref{eq:A1A3}\eqref{eq:A1A32} \\
(A_1,B_3) & 5/4 & \mathrm{SM}(8,6)  & \eqref{eq:A1B3}\text{-}\eqref{eq:A1B33} \\
\hline
\end{array}
\]
\bigskip
\caption{Pair of Dynkin diagrams, (generalized) Nahm sums and associated 2d CFTs.}
\label{tb:full}
\end{table}

Here we explain two examples in detail; other cases will be discussed in Section \ref{sec:2dCFTs}. We first consider the pair $(T_2,A_1)$. The matrix $A(T_2,A_1)=\begin{psmallmatrix}
1/2 & -1/2 \\ -1/2 & 1    
\end{psmallmatrix}$ is included in the Zagier list \cite[Table 2]{Zag07}. There are three Nahm sums, and we label them by $f_j$ for $1\leq j\leq 3$ in the order of Zagier's list. Notice that $c(T_2,A_1)= {10}/{7}$, which equals the central charge of the effective $N=1$ supersymmetric Virasoro minimal model $\mathrm{SM}_{\rm eff}(84,2)$. There are 21 {\rm NS} weights and 21 {\rm R} weights. We can express the three Nahm sums in terms of the characters of $\mathrm{SM}_{\rm eff}(84,2)$ as follows
\begin{align}
f_1 & = \chi_{{\rm NS},0}+\chi_{{\rm NS},\frac12}+\chi_{{\rm NS},\frac52}+\chi_{{\rm NS},5}+2\chi_{{\rm R},\frac14  }+2\chi_{{\rm R},\frac54  }, \label{eq:T2A1-1} \\  
f_2 &=\chi_{{\rm NS},\frac{5}{14}}+\chi_{{\rm NS}, \frac67}+\chi_{{\rm NS},\frac{13}{7}}-\chi_{{\rm NS},\frac{20}{7} }+2\chi_{{\rm R},\frac{3}{28}  } +2\chi_{{\rm R},\frac{87}{28}  },\\
f_3 &= \chi_{{\rm NS},\frac{1}{14}}-\chi_{{\rm NS},\frac{15}{14}}+\chi_{{\rm NS},\frac{11}{7}}+\chi_{{\rm NS},\frac{57}{14}}+2\chi_{{\rm R},\frac{23}{28}  }-2\chi_{{\rm R},\frac{135}{28}} \label{eq:T2A1-3},
\end{align}
where the subscripts indicate the conformal weights. 

We then consider the infinite family of pairs $(T_1,D_r)$ with $r\geq 3$. In this case, $A(T_1,D_r)=C(D_r)^{-1}$, and the Nahm sums correspond to a theory of central charge $3r/(2r+1)$. We find that the theory is the effective $N=1$ supersymmetric Virasoro minimal model $\mathrm{SM}_{\mathrm{eff}}(8r+4,2)$. Note that $\mathrm{SM}_{\mathrm{eff}}(8r+4,2)$ has $2r+1$ NS characters and $2r+1$ R characters. We use $j=1,2,\dots, 2r+1$ to label them from small to large in weight. 
We find the following uniform relation between the Nahm sum and the fermionic characters:
\begin{equation}\label{eq:T1Dr}
f_{A,\vec{0},-\frac{r}{8(2r+1)}}=\chi^{\mathrm{SM}_{\rm eff}(8r+4,2) }_{{\rm NS},\, j=1}+\chi^{\mathrm{SM}_{\rm eff}(8r+4,2) }_{{\rm NS},\, j=2r+1}+2\chi^{\mathrm{SM}_{\rm eff}(8r+4,2) }_{{\rm R},\, j=r+1}.
\end{equation}
From the well-known product expansions of these characters, we obtain the following generalized Rogers--Ramanujan identities
\begin{equation}
\begin{split}
f_{C(D_r)^{-1},\vec{0},0}(\tau)=&\prod_{\substack{n=1\\ n\not\equiv 2 \bmod 4\\ n\not\equiv 0,\, \pm (4r+1) \bmod (8r+4)}}^\infty \big(1-q^{n/2}\big)^{-1} + q^{\frac{r}{2}} \cdot \prod_{\substack{n=1\\ n\not\equiv 2 \bmod 4\\ n\not\equiv 0,\, \pm 1 \bmod (8r+4)}}^\infty \big(1-q^{n/2}\big)^{-1} \\
&+ 2q^{\frac{r}{8}}\cdot \prod_{\substack{n=1\\ \text{$n$ odd}}}^\infty \big(1-q^{n}\big)^{-1}  \prod_{\substack{n=1\\ n\not\equiv 0,\, \pm (r+1) \bmod (4r+2)}}^\infty \big(1-q^{n}\big)^{-1}. 
\end{split}
\end{equation} 
We also conjecture that there exist $r-1$ non-zero vectors $B$ such that the associated Nahm sums are modular and have similar expressions in terms of NS and R characters of $\mathrm{SM}_{\rm eff}(8r+4,2)$.

\bigskip
\noindent
\textbf{Acknowledgements} 
H. Wang was supported by the National Key R$\&$D Program of China (Grant No. 2024YFA1014500). K. Sun is supported by China NSF grant No.1250010384. The authors thank Liuquan Wang and Don Zagier for helpful discussions.

\section{Known cases of the Dynkin diagram conjecture}\label{sec:known-cases}
In this section, we elaborate what is known about Conjecture \ref{Conj:main}. First, this conjecture holds for several infinite families that correspond to known combinatorial identities. 

\begin{enumerate}
\item The Nahm sum associated with $(A_1, T_r)$ is modular for any $r\geq 1$ by the case $s=r+1$ of the famous Andrews--Gordon identity \cite{Gor61,And74}:
\begin{equation}
\sum_{n_1,...,n_r\geq 0} \frac{q^{N_1^2+N_2^2+\cdots +N_r^2+N_s+N_{s+1}+\cdots +N_r}}{(q;q)_{n_1}(q;q)_{n_2}\cdots (q;q)_{n_r}}=\frac{(q^s,q^{2r+3-s},q^{2r+3};q^{2r+3})}{(q;q)_\infty},    
\end{equation}
where $1\leq s\leq r+1$ and $N_j=n_j+n_{j+1}+\cdots+n_r$ for $1\leq 1\leq r$ and $N_{r+1}=0$. The functions defined by the Andrews--Gordon identity are in fact the $r+1$ characters of the Virasoro minimal model $\mathrm{M}(2r+3,2)$ with effective central charge
$$
c(A_1,T_r)=\frac{2r}{2r+3}. 
$$

\vspace{3mm} 

\item The Nahm sum associated with $(T_1, T_{2r})$ is modular for any $r\geq 1$ by the identity:
\begin{equation}\label{eq:War}
 \sum_{n_1,...,n_{2r}\geq 0} \frac{q^{\frac{1}{2}(N_1^2+N_2^2+\cdots +N_{2r}^2)}}{(q;q)_{n_1}(q;q)_{n_2}\cdots (q;q)_{n_{2r}}}=\prod_{\substack{j=1\\ j\neq 2 \; \mathrm{mod}\; 4\\
 j\neq 0, \pm (2r+1) \; \mathrm{mod} \; 4(r+1)}}^\infty \frac{1}{(1-q^{j/2})},
\end{equation}
where $N_j$ is defined as above. 
This identity and some relevant identities were first conjectured by Melzer \cite[Equations (1.7), (2.1)-(2.7)]{Mel94}. Melzer also observed that these Nahm sums give the $r+1$ NS characters and the $r+1$ R characters of the level $r$ supersymmetric Lee--Yang models $\mathrm{SM}(4r+4,2)$ with effective central charge
$$
c(T_1,T_{2r})=\frac{3r}{2(r+1)}.
$$
We therefore view these identities as supersymmetric analogs of the Andrews--Gordon identity. Their proofs can be found in \cite[Theorem 5.1]{BIS00} and \cite[Theorem 4.4]{War03}. 

\vspace{3mm}

\item The Nahm sum associated with $(T_1, T_{2r-1})$ is modular for any $r\geq 1$ by Warnaar's identity  \cite[Theorem 4.5]{War03}:
\begin{equation}\label{eq:War-odd}
 \sum_{n_1,...,n_{2r-1}\geq 0} \frac{q^{\frac{1}{2}(N_1^2+N_2^2+\cdots +N_{2r-1}^2)}}{(q;q)_{n_1}(q;q)_{n_2}\cdots (q;q)_{n_{2r-1}}}=(-q^{1/2})_\infty \prod_{\substack{j=1\\ 
 j\neq 0, \pm r \; \mathrm{mod} \; (2r+1)}}^\infty \frac{1}{(1-q^j)}.
\end{equation}
The Nahm sum associated with $(T_1, T_1)$ is the vacuum character of free fermion $F$, namely the fermionization of the 2d critical Ising model $\mathrm{M}(4,3)$. The Nahm sum associated with $(T_1, T_{2r+1})$ for $r\geq 1$ is the vacuum character of the rational 2d CFT $F\otimes \mathrm{M}_{\rm eff}(2r+3,2)$ with central charge
$$
c(T_1, T_{2r+1})=\frac{1}{2} + \frac{2r}{2r+3}. 
$$
This can be easily recognized from the right hand side of \eqref{eq:War-odd}.

\vspace{3mm}

\item The generalized Nahm sum associated with $(A_1, B_r)$ is modular for any positive integer $r\geq 2$ by \cite[Theorem 3.4]{BKRS23}:  
\begin{equation}
 \sum_{n_1,...,n_r\geq 0} \frac{q^{\frac{1}{2}(M_1^2+\cdots +M_r^2)}}{(q^2;q^2)_{n_1}\cdots (q^2;q^2)_{n_{r-1}}(q;q)_{n_r}}=\frac{(-q^{r/2},-q^{r/2+1},q^{r+1};q^{r+1})_\infty}{(q^2;q^2)_\infty},   
\end{equation}
where $M_i=2n_i+2n_{i+1}+\cdots +2n_{r-1}+n_r$ for $1\leq i<r$ and $M_r=n_r$. In this case, $D=\mathrm{diag}(2,...,2,1)$ and the central charge is
$$
c(A_1,B_r)= \frac{2r-1}{r+1}.
$$
We do not know what the corresponding 2d CFT is for general $r$. Nevertheless, for $r=2,3$, we do find the corresponding 2d CFT which will be discussed in Section \ref{sec:2dCFTs}.

\vspace{3mm}

\item The generalized Nahm sum associated with $(A_1, C_r)$ is modular for any $r\geq 2$ by the case $s=r+1$ of Bressoud's identity  \cite{Bre79}:
\begin{equation}\label{eq:AC}
\sum_{n_1,...,n_r\geq 0} \frac{q^{N_1^2+N_2^2+\cdots +N_r^2+N_s+N_{s+1}+\cdots +N_r}}{(q;q)_{n_1}\cdots (q;q)_{n_{r-1}}(q^2;q^2)_{n_r}}=\frac{(q^s,q^{2r+2-s},q^{2r+2};q^{2r+2})}{(q;q)_\infty}.   
\end{equation}
Indeed, $D(A_1,C_r)=D(C_r)=\mathrm{diag}(1,...,1,2)$ and 
$$
A(A_1,C_r)D(A_1,C_r)=2C(T_r)^{-1}=C(A_1)\otimes C(T_r)^{-1}=A(A_1,T_r).
$$
The central charge $c(A_1,C_r)$ is always $1$. We find that the corresponding 2d CFT is the circle compact boson $U(1)_{4(r+1)}$; see Section \ref{sec:2dCFTs} for details. 

\vspace{3mm}

\item The Nahm sum associated with $(A_1,D_r)$ is modular for any $r\geq 3$. In 1993, Kedem, Klassen, McCoy, and Melzer \cite[Section 2]{KKMM93a} conjectured the following identities
\begin{align}
\sum_{\substack{n=(n_1,...,n_r)\in\NN^r\\ n_{r-1}+n_r \; \text{even}}}   \frac{q^{n^tC(D_r)^{-1}n}}{(q;q)_{n_1}\cdots(q;q)_{n_r}} &= (q;q)^{-1}_\infty \sum_{k\in\ZZ}q^{rk^2},\\
\sum_{\substack{n=(n_1,...,n_r)\in\NN^r\\ n_{r-1}+n_r \; \text{odd}}}   \frac{q^{n^tC(D_r)^{-1}n}}{(q;q)_{n_1}\cdots(q;q)_{n_r}} &= (q;q)^{-1}_\infty \sum_{k\in\ZZ}q^{r(k+1/2)^2},
\end{align}
which were proved by Warnaar \cite{War07}. The corresponding 2d CFT is circle compact boson $U(1)_r$. 

\vspace{3mm}

\item The generalized Nahm sum associated with $(T_1, C_r)$ is modular for any $r\geq 2$ by Warnaar's identity \cite[Equation (5.14)]{War03}:
\begin{equation}\label{eq:TC}
\sum_{n_1,...,n_r\geq 0} \frac{q^{\frac{1}{2}(N_1^2+N_2^2+\cdots +N_r^2)}}{(q;q)_{n_1}\cdots (q;q)_{n_{r-1}}(q^2;q^2)_{n_r}}=\frac{(q^{r/2+1/2},q^{r/2+1},q^{r+3/2};q^{r+3/2})}{(-q)_\infty (q^{1/2};q^{1/2})_\infty}.   
\end{equation}
The central charge is
$$
c(T_1,C_r)=\frac{3(r+1)}{2r+3}. 
$$
We will describe the corresponding 2d CFT in Section \ref{sec:2dCFTs}.

\vspace{3mm}

\item The Nahm sum associated with $(A_{2r-1}, T_k)$ is modular for any $r\geq 1$ and $k\geq 1$ according to the recent work \cite[Theorem 1.0.1]{CG25} by Creutzig and Garner.  By means of a deformation of the affine vertex operator superalgebra $L_{k}(\mathfrak{osp}_{1|2r})$ into the principal subsuperspace of $L_k(\mathfrak{sl}_{1|2r+1})$, they identified the Nahm sum associated with $(A_{2r-1}, T_k)$ with the vacuum supercharacter of the affine VOSA $L_{k}(\mathfrak{osp}_{1|2r})$. Note that the central charge is
$$
c(A_{2r-1},T_k)=\frac{2kr(2r-1)}{2(k+r)+1}.
$$
The corresponding $q$-series identities were previously proved for $k=1$ and conjectured to hold more generally by Warnaar and Zudilin \cite[Conjecture 1.1 and Theorem 1.2]{WZ12}. All supercharacters of $L_k(\mathfrak{osp}_{1|2r})$ form a vector-valued modular function for a certain finite-dimensional complex representation of $\SL_2(\ZZ)$, so it is natural to expect that each non-vacuum supercharacter also has a Nahm sum expression. Note that the supercharacters of $L_k(\mathfrak{osp}_{1|2})$ and $\mathrm{M}(2r+3,2)$ are the same as the $q$-series \cite{Ferrari:2023fez}. Thus, the particular case of $r=1$ here is consistent with Case (1) discussed above. 
\end{enumerate}

\vspace{3mm}

The representation of infinite-dimensional Lie algebras also plays an important role in proving the modularity of Nahm sums. More precisely, Nahm sums can be identified with graded dimensions of principal subspaces of standard modules for (twisted) affine Kac--Moody algebras in some cases \cite{LM78a, LM78b, LW82, LW84, GOW16}. We review two results. Combining \cite{FS94, Sto98, WZ12, CG25} and \cite[Theorem 1.1]{BCFK14} together, the Nahm sums associated with $(A_r,T_k)$ are modular functions for all positive integers $r$ and $k$. By \cite[Proposition 4.0.1]{Dun15}, the Nahm sums associated with $(A_{n-1},A_{k-1})$ are also modular for all integers $n,\, k\geq 2$ with $\mathrm{gcd}\big( (n-1)!,\, k \big)=1$.  

\vspace{3mm}

We now consider isolated examples of low rank. Conjecture \ref{Conj:main} produces $35$ generalized Nahm sums of rank less than $4$. The 28 of them are known to be modular. We recall these cases. 
\begin{enumerate}
\item[(a)] The eight infinite families above produce the following $13$ cases:
\begin{align*}
&(A_1,T_1)& &(A_1, T_2)& &(A_1, T_3)& &(A_1, A_3)& &(A_1,B_2)& &(A_1,B_3)& &(A_1,C_3)& \\
&(T_1,T_1)& &(T_1, T_2)& &(T_1, T_3)& &(T_1, C_2)& &(T_1,C_3)& &(A_3, T_1).&
\end{align*}
Note that $A_3=D_3$ and $B_2=C_2$. 

\item[(b)] The representation of affine Kac--Moody algebras yields cases $(A_2,T_1)$ and $(A_1,A_2)$. 

\item[(c)] The case $(T_1,A_1)$ was solved by Zagier \cite{Zag07}. In fact, he \cite{Zag07} showed that there are exactly $7$ modular triples corresponding to three distinct matrices $A$, which are given by $(A_1,T_1)$, $(T_1,T_1)$, and $(T_1,A_1)$. This is the only known classification of modular Nahm sums. 

\item[(d)] The cases $(T_1,A_2)$, $(T_2,T_1)$, $(T_2,A_1)$, and $(A_2,A_1)$ have been verified. Zagier found $11$ sets of possible modular triples of rank two and formulated them as \cite[Table 2]{Zag07}, which include all $8$ cases resulting from Conjecture \ref{Conj:main}. Thanks to \cite{Zag07, VZ11, CF13, Wan22a, CRW24}, the modularity of all Zagier's examples has been proven. In particular, the cases $(T_1,A_2)$ (and also $(A_2,T_1)$) correspond to Example 1 of Zagier's table by setting $\alpha=2/3$ (resp. $\alpha=2$), and thus the modularity has been verified by Zagier.  The cases $(T_2,T_1)$ and $(T_2,A_1)$  were proved by \cite{Wan22a}, and the case $(A_2,A_1)$ was verified by \cite{VZ11}. Note that the case $(A_1,A_2)$ in (b) above was first justified by \cite{CF13}. 

\item[(e)] The cases $(A_1,G_2)$, $(G_2,A_1)$, $(C_2,A_1)$ and $(C_2,T_1)$ are included in Table 1 of Mizuno's paper \cite{Miz23}. The modularity of these generalized Nahm sums has been verified by \cite{WW23}. 

\item[(f)] The case $(T_1, A_3)$ is trivial in some sense. By Zwegers's unpublished duality between Nahm sums of rank 2 and rank 3 (see \cite[Theorem 3.6.4]{Lee12} or \cite[Proposition 2.1]{CRW24}), the Nahm sum associated with $(T_1, A_3)$ defines the same function as Example 9 of Zagier's Table 2, for which the modularity was proved by \cite{Wan22a}. Similarly, the case $(A_1,A_3)$ in (a) above corresponds to Example 1 of Zagier's Table 2 with $\alpha=3/2$, so its modularity is also trivial. Remark that the cases $(A_1,A_3)$, $(T_1,A_3)$ and their dual are not included in Zagier's table of rank-three Nahm sums \cite{Zag07}. The modularity of all Zagier's rank-three examples has been verified by \cite{Wan22b}. 

\item[(g)] The case $(T_3,T_1)$ was proved by \cite{MW24}, in which the conjectural identities for the Nahm sums associated with $(T_r,T_1)$ were also formulated.  

\item[(h)] The case $(T_1,B_3)$ was proved by \cite{WW24b}. This case and the case $(A_1,B_3)$ are Example 7 and Example 5 in \cite[Table 3]{Miz23}, respectively. The modularity of all generalized Nahm sums in \cite[Table 3]{Miz23} was justified by \cite{WW24b}. In addition, the modularity of all generalized Nahm sums in \cite[Table 2]{Miz23} was verified by \cite{WW24}, which includes the cases $(A_1,C_3)$ and $(T_1,C_3)$. 

\item[(i)] The case $(A_3,A_1)$ was proved by \cite[Equation (6.34)]{CW24}.
\end{enumerate}

\vspace{2mm}

Through the long discussion above, among the $35$ generalized Nahm sums of rank at most $3$ resulting from Conjecture \ref{Conj:main}, it remains to verify the modularity of the generalized Nahm sums associated with the following seven pairs:
\begin{align*}
&(T_1,G_2)& &(G_2,T_1)& &(T_3,A_1)& &(C_3, A_1)& \\
&(C_3, T_1)& &(B_3,A_1)& &(B_3,T_1).&     
\end{align*}
The modularity for $(T_1, G_2)$ can be deduced from the conjectural identity
$$
\sum_{i,j\geq 0} \frac{q^{i^2+3ij+3j^2}}{(q;q)_i(q^3;q^3)_j}=\frac{1}{(q,q^3,q^6,q^8;q^9)_\infty},
$$
which is a conjecture of Kanade--Russell \cite{KR15}. The modularity for $(G_2, T_1)$ can be derived from 
$$
\sum_{i,j\geq 0} \frac{q^{i^2-3ij+3j^2}}{(q;q)_i(q^3;q^3)_j}=\frac{1}{(q,q^3,q^6,q^8;q^9)_\infty^2}+q\frac{1}{(q^3,q^6;q^9)_\infty^2(q^2,q^4,q^5,q^7;q^9)_\infty},
$$
which is a conjecture of Wang--Wang \cite[Conjecture 3.6]{WW23}.

\section{Generalized Nahm sums and characters of 2d CFTs}\label{sec:2dCFTs}

In this section, we discuss various cases of (generalized) Nahm sums of type $(X,Y)$ and their relation to 2d CFT characters. We find these relations by checking the $q$-series to very high orders. 
We summarize the  (generalized) Nahm sum from the pair of Dynkin diagrams and the associated 2d CFT in Table \ref{tb:full}. The cases $(A_1,Y)$ and $(T_1,Y)$ are of special interest. We first discuss these two cases and then move to isolated cases.

In \cite{KKMM93a} some conjectural correspondences between 2d CFTs and Nahm sums of type $(A_1,Y)$ were proposed, where $Y$ is a simply-laced Dynkin diagram. There it was suggested that for this type of Nahm sums, the associated 2d CFTs are precisely the coset
\begin{align}
    \frac{(Y)_1\times (Y)_1}{(Y)_2}.
\end{align}
For example, when $Y=E_8$, the coset CFT is well-known to be the Ising model $\mathrm{M}(4,3)$, and one has the identity \cite{KKMM93a, WP94}:
\begin{equation}
\sum_{n=(n_1,...,n_8)\in\NN^8} \frac{q^{n^t C(E_8)^{-1}n}}{(q;q)_{n_1}\cdots (q;q)_{n_8}} =  \sum_{m\in\NN} \frac{q^{2m^2}}{(q;q)_{2m}}.   
\end{equation}

When $Y$ is not simply-laced, the associated 2d CFT cannot be the coset $\big({(Y)_1\times (Y)_1}\big)/{(Y)_2}$. Nevertheless, we observe that the 2d CFT for the non-simply-laced case is closely related to the one for the simply-laced case before folding. In particular, if $G=ADE$ and $G'$ is a folding of $G$, then we notice
\begin{equation}
    c(A_1,G')=c(A_1,G).
\end{equation}
This folding is different from the twist in twisted affine Lie algebras. The twist does not change the dual Coxeter number; here the folding does not change the Coxeter number. Folding can be considered as the combination of twist and exchange of long and short roots. Then from \eqref{eq:cXY}, it is easy to see that the central charge remains the same upon folding. We further conjecture that the CFT associated with the folded one can be conformally embedded in the CFT associated with the unfolded one:

\begin{conjecture}\label{conj:A1}
\begin{align}
 \mathrm{\bf CFT}(A_1,G')\subset   \mathrm{\bf CFT}(A_1,G)  . 
\end{align}
\end{conjecture}

This suggests that the characters of $\mathrm{\bf CFT}(A_1,G)$ can be written as linear combinations of the characters of $\mathrm{\bf CFT}(A_1,G')$. In particular, the Nahm sum of type $(A_1,G)$ can be expressed by the characters of $\mathrm{\bf CFT}(A_1,G')$. This also suggests that the partition function of $\mathrm{\bf CFT}(A_1,G)$ is a non-diagonal modular invariant of $\mathrm{\bf CFT}(A_1,G')$. 
We have verified that Conjecture \ref{conj:A1} is true for the following $\{G,G'\}$:
\begin{itemize}
    \item $\{E_6,F_4\}$
    \item $\{D_4,G_2\}$
    \item $\{D_r,C_{r-1}\}$
\end{itemize}
Currently we could not verify the cases $\{A_{2r-1},B_r\}$ for general $r$, due to the lack of CFT understanding for the type $(A_1,B_r)$. Nevertheless, for $r=2,3$, we can determine $\mathrm{\bf CFT}(A_1,B_r)$ and confirm the conjecture. We will present the check for $r=3$ in the end of Section \ref{sec:2dCFTs}.

In the remainder of this section, we discuss the relation between generalized Nahm sums of type $(X,Y)$ and characters of 2d CFTs case by case. 

\vspace{3mm}

\textbf{Case $(A_1, C_r)$}. We find that the associated 2d CFT is $U(1)_{4(r+1)}$. The central charge is always 1. The Nahm sum has the expression
\begin{align}\label{eq:A1Cr}
    f_{A,\vec{0},-1/24}=\chi_0^{U(1)_{4(r+1)}}-\chi_{r+1}^{U(1)_{4(r+1)}},
\end{align}
where the subscript is the conformal weight. This equality follows from \eqref{eq:AC} and Jacobi's triple product formula. 

\vspace{3mm}

\textbf{Case $(T_1, C_r)$}. We find that the associated 2d CFT is $\mathrm{SM}_{\rm eff}(4r+6,4)$. The central charge is $c={3(r+1)}/{(2r+3)}$. The generalized Nahm sums are related to the NS characters by
\begin{align}\label{eq:T1Cr}
f_{A,\vec{0},-c/24}= \left\{  
             \begin{array}{ll}  
             \chi^{\rm NS}_0-\chi^{\rm NS}_{n+1},\ & \text{if $r=2n$}, \\[+2mm]  
             \chi^{\rm NS}_0-\chi^{\rm NS}_{n+1},\ \ & \text{if $r=2n+1$}.   
             \end{array}  
\right.  
\end{align}
Here, $n$ is the largest integer no greater than $r/2$, and the subscripts are the NS conformal weights. We have checked this expression up to $r=7$. These relations can also be extended to the cases with nonzero $B$ vectors. For example, for $r=2$, $A=\begin{psmallmatrix}
1 & \frac{1}{2} \\ 1 & 1    
\end{psmallmatrix}$. The associated 2d CFT $\mathrm{SM}_{\rm eff}(14,4)$ has effective central charge ${9}/{7}$ and effective NS conformal weights
$0,\frac{3}{112},\frac{1}{14},\frac{3}{14},\frac{5}{16},\frac{3}{7},\frac{99}{112},\frac{15}{14},2,\frac{45}{14}$. For this $A$ matrix, Mizuno \cite{Miz23} gave three triples of $(B,C)$, which we find to be related to the NS characters of $\mathrm{SM}_{\rm eff}(14,4)$ by
\begin{align}
    f_{A,(0,0),-\frac{3}{56}}&=\chi_{0}^{\rm NS}-\chi_{2}^{\rm NS},\\
    f_{A,(0,1),\frac{1}{56}}&=\chi_{\frac{1}{14}}^{\rm NS}-\chi_{\frac{15}{14}}^{\rm NS},\\
    f_{A,(1,1),\frac{9}{56}}&=\chi_{\frac{3}{14}}^{\rm NS}-\chi_{\frac{45}{14}}^{\rm NS}.
\end{align} 
For $r=3$, the generalized Nahm sum associated with $(T_1,C_3)$ has $$
A=\left(
\begin{array}{ccc}
 1 & 1 & \frac{1}{2} \\
 1 & 2 & 1 \\
 1 & 2 & \frac{3}{2} \\
\end{array}
\right). 
$$ 
The associated 2d CFT $\mathrm{SM}_{\rm eff}(18,4)$ has effective central charge $4/3$. For this $A$ matrix, Mizuno \cite{Miz23} gave four triples of $(B,C)$, which we find to be related to the NS characters of $\mathrm{SM}_{\rm eff}(18,4)$ by
\begin{align}
    f_{A,(0,0,0),-\frac{1}{18}}&=\chi_{0}^{\rm NS}-\chi_{2}^{\rm NS},\\
    f_{A,(0,0,1),0}&=\chi_{\frac{1}{18}}^{\rm NS}-\chi_{\frac{55}{18}}^{\rm NS},\\
    f_{A,(0,1,1),\frac{1}{9}}&=\chi_{\frac{1}{6}}^{\rm NS}-\chi_{\frac{7}{6}}^{\rm NS},\\
    f_{A,(1,1,2),\frac{5}{18}}&=\chi_{\frac{1}{3}}^{\rm NS}-\chi_{\frac{13}{3}}^{\rm NS}.
\end{align}

\vspace{3mm}

\textbf{Case $(A_1, F_4)$}. The generalized Nahm sum associated with $(A_1,F_4)$ is a modular function according to \cite[Equation (1.7)]{BKRS23}:
\begin{equation}\label{eq:A1F4}
\sum_{i,j,k,l\geq 0} \frac{q^{4i^2+12ij+8ik+4il+12j^2+16jk+8jl+6k^2+6kl+2l^2}}{(q^2;q^2)_i(q^2;q^2)_j(q;q)_k(q;q)_l}  = \frac{1}{(q^2,q^3,q^4,q^{10},q^{11},q^{12};q^{14})_\infty}. 
\end{equation}
Note that $D(A_1,F_4)=\mathrm{diag}(2,2,1,1)$, and the central charge is $c(A_1,F_4)=6/7$. We find that the associated 2d CFT is the minimal model $\mathrm{M}(7,6)$, and the Nahm sum \eqref{eq:A1F4} is equal to 
\begin{align}\label{eq:A1F4b}
    q^{\frac{1}{28}}\left(\chi_0^{\mathrm{M}(7,6)}-\chi_5^{\mathrm{M}(7,6)}\right).
\end{align}
Here, the subscript is the conformal weight.

\vspace{3mm}

\textbf{Case $(T_1, F_4)$}.
The generalized Nahm sum associated with $(T_1,F_4)$ is a modular function according to \cite[Theorem 1.5]{BKRS23}:
\begin{equation}\label{eq:T1F4}
\sum_{i,j,k,l\geq 0} \frac{q^{2i^2+6ij+4ik+2il+6j^2+8jk+4jl+3k^2+3kl+l^2}}{(q^2;q^2)_i(q^2;q^2)_j(q;q)_k(q;q)_l}  = \frac{1}{(q,q^2,q^4,q^{6},q^{8},q^{9};q^{10})_\infty}. 
\end{equation}
Note that $D(T_1,F_4)=\mathrm{diag}(2,2,1,1)$, and the central charge is $c(T_1,F_4)=6/5$. We find that the associated 2d CFT is the product of two minimal models $\mathrm{M}_{\rm eff}(5,2)\otimes \mathrm{M}(6,5)$. We checked that the Nahm sum \eqref{eq:T1F4} is equal to 
\begin{align}\label{eq:T1F4b}
    q^{\frac{1}{20}}\chi_0^{\mathrm{M}_{\rm eff}(5,2)}\left(\chi_0^{\mathrm{M}(6,5)}-\chi_3^{\mathrm{M}(6,5)}\right).
\end{align} 

\vspace{3mm}

\textbf{Case $(A_1, G_2)$}.
By \cite{WW23}, the generalized Nahm sum associated with $(A_1,G_2)$ is a modular function. Note that $D(A_1,G_2)=\mathrm{diag}(1,3)$, and the central charge is $c(A_1,G_2)= 1$. We find that the associated 2d CFT is $U(1)_{36}$. We checked that the generalized Nahm sum is equal to 
\begin{align}\label{eq:A1G2}
    f_{A,\vec{0},-\frac{1}{24}}= \chi^{U(1)_{36}}_0-\chi^{U(1)_{36}}_{1}-\chi^{U(1)_{36}}_{4}+\chi^{U(1)_{36}}_{9}.
\end{align}
As before, the subscript is the conformal weight. 

\vspace{3mm}

\textbf{Case $(T_1, E_6)$}.
The Nahm sum associated with $(T_1,E_6)$ is of central charge 6/5. We find that the 2d CFT is $\mathrm{M}_{\rm eff}(5,2)\otimes \mathrm{M}_{\rm sub}(6,5)$. We checked that the Nahm sum is equal to 
\begin{align}\label{eq:T1E6}
  f_{A,\vec{0},-\frac{1}{20}}=  \chi_0^{\mathrm{M}_{\rm eff}(5,2)}\left(\chi_0^{\mathrm{M}(6,5)}+\chi_3^{\mathrm{M}(6,5)}+2\chi_{2/3}^{\mathrm{M} (6,5)}\right). 
\end{align}
Note that the right hand side has the typical combination of three-state Potts model which is a non-diagonal modular invariant of $\mathrm{M} (6,5)$, denoted as $\mathrm{M}_{\rm sub}(6,5)$ in \cite[Section 5.5]{Duan:2022ltz}. 

\vspace{3mm}

\textbf{Case $(T_1, E_8)$ and $(E_8, T_1)$}.
The Nahm sum associated with $(T_1,E_8)$ is of central charge 8/11. We find that the associated 2d CFT is the effective minimal model $\mathrm{M}_{\rm eff}(11,2)$. The Nahm sum equals the vacuum character: 
\begin{align}\label{eq:T1E8}
    f_{A,\vec{0},-1/33}=\chi_0^{\mathrm{M}_{\rm eff}(11,2)}.
\end{align}

We also find that the Nahm sum associated with $(E_8,T_1)$ can be expressed by the vacuum character of $\mathsf{T}_{10}\mathrm{M}_{\rm eff}(11,2)$ as
\begin{align}\label{eq:E8T1}
    f_{A,\vec{0},-10/33}=\chi_0^{\mathsf{T}_{10} \mathrm{M}_{\rm eff}(11,2)}.
\end{align}
Here, $\mathsf{T}_{10}\mathrm{M}_{\rm eff}(11,2)$ is the image of $\mathrm{M}_{\rm eff}(11,2)$ under the action of the Hecke operator $\mathsf{T}_{10}$, which was studied in \cite{Duan:2022ltz}. It can also be realized as a coset
\begin{align}\nonumber
\mathsf{T}_{10}\mathrm{M}_{\rm eff}(11,2)=\frac{(E_8)_1}{\mathrm{M}_{\rm eff}(11,2)}.
\end{align}
We record the first few coefficients of \eqref{eq:E8T1} as $1+120 q+1660 q^2+12320 q^3+68210 q^4+O(q^5)$. 
On the other hand, for Nahm sums of type $(E_6,T_1)$ and $(E_7,T_1)$, we have not found good CFT interpretations.

\vspace{3mm}

\textbf{Case $(T_1, A_1)$}.  This is a well-known case of rank one with $A=1/2$ and $c=3/5$. The associated CFT is the effective minimal model $\mathrm{M}_{\rm eff}(5,3)$. There exist two modular Nahm sums for this $A$. It is easy to find that the two Nahm sums are related to the CFT characters by
\begin{align}\label{eq:T1A11}
  f_{1/2,0,-1/40}  &= \chi_0^{\mathrm{M}_{\rm eff}(5,3)} + \chi_{1/4}^{\mathrm{M}_{\rm eff}(5,3)},\\\label{eq:T1A12}
  f_{1/2,1/2,1/40}  &= \chi_{1/20}^{\mathrm{M}_{\rm eff}(5,3)} + \chi_{4/5}^{\mathrm{M}_{\rm eff}(5,3)}.
\end{align}

\vspace{3mm}

\textbf{Case $(T_1, A_2)$}. 
This is Example 1 in Zagier's list \cite{Zag07} of rank two with $\alpha=2/3$. We find that the associated CFT is $U(1)_3$. The Nahm sum can be expressed by the $U(1)_3$ characters as 
\begin{align}\label{eq:T1A2}
f_{A,\vec{0},-\frac{1}{24}}=\chi_0^{U(1)_3} +2\chi_{1/3}^{U(1)_3}.
\end{align}

\vspace{3mm}

\textbf{Case $(A_2, T_1)$}. This is Example 1 in Zagier's list \cite{Zag07} of rank two with $\alpha=2$. In this case, $A=\begin{psmallmatrix}
2 & -1 \\ -1 & 2    
\end{psmallmatrix}$ and $c=1$. We find that the associated 2d CFT is $(A_1)_1$, or equivalently $U(1)_4$. The Nahm sums and the CFT characters satisfy the following relations:
\begin{align}\label{eq:A2T1}
f_{A,\vec{0},-\frac{1}{24}}=\chi_0^{(A_1)_1}=\chi_0^{U(1)_4} +\chi_1^{U(1)_4} .
\end{align}
As before, the subscript is the conformal weight. 

\vspace{3mm}

\textbf{Case $(T_2, T_1)$}. 
This is Example 3 in Zagier's list \cite{Zag07} of rank two. In this case, $A=\begin{psmallmatrix}
1 & -1 \\ -1 & 2    
\end{psmallmatrix}$ and $c=5/4$. There are five Nahm sums, and we label them by $f_j$ for $1\leq j\leq 5$ in the order of Zagier's list. Here we correct a typo of \cite{Zag07}, for $f_5$, $C=\frac{19}{96}$, not $\frac{19}{24}$. The associated 2d CFT is $\mathrm{SM}_{\rm eff}(8,2)\otimes F$. We find the following relations between the Nahm sums and the CFT characters:
\begin{align}\label{eq:T2T11}
    f_2 & =\chi_{{\rm NS},0}^{\mathrm{SM}_{\rm eff}(8,2) }\chi_{{\rm NS},0}^{F},\\\label{eq:T2T12}
    f_3 &= \chi_{{\rm R},{1}/{32}}^{\mathrm{SM}_{\rm eff}(8,2) }\chi_{{\rm R},{1}/{16}}^{F}  ,\\ \label{eq:T2T13}
    f_4 &= \chi_{{\rm R},{5}/{32}}^{\mathrm{SM}_{\rm eff}(8,2) }\chi_{{\rm R},{1}/{16}}^{F}  ,\\ \label{eq:T2T14}
    f_5 &= \chi_{{\rm NS},1/4}^{\mathrm{SM}_{\rm eff}(8,2) }\chi_{{\rm NS},0}^{F}.
\end{align}
Note that $f_1=2f_4$ and the Nahm sum with $B=0$ is $f_2$.

\vspace{3mm}

\textbf{Case $(T_1, A_3)$}. As mentioned in Case (f) of Section \ref{sec:known-cases}, this case reduces to Example 9 in Zagier's list \cite{Zag07} of rank two, in which $A=\begin{psmallmatrix}
1 & -1/2 \\ -1/2 & 3/4    
\end{psmallmatrix}$ and $c=9/7$. There are three Nahm sums, and we label them by $f_j$ for $1\leq j\leq 3$ in the order of Zagier's list. We find that the associated 2d CFT is $\mathrm{SM}_{\rm eff}(28,2)$ and these Nahm sums are delicate mixes of the NS and R characters of $\mathrm{SM}_{\rm eff}(28,2)$. More precisely,  
\begin{align} \nonumber
f_1 & = \chi^{\mathrm{SM}_{\rm eff}(28,2) }_{{\rm NS},3/14}+\chi^{\mathrm{SM}_{\rm eff}(28,2) }_{{\rm NS},5/7}+2\chi^{\mathrm{SM}_{\rm eff}(28,2) }_{{\rm R}, 5/56},\\   \nonumber
    f_2 &= \chi^{\mathrm{SM}_{\rm eff}(28,2) }_{{\rm NS},0}+\chi^{\mathrm{SM}_{\rm eff}(28,2) }_{{\rm NS},3/2}+2\chi^{\mathrm{SM}_{\rm eff}(28,2) }_{{\rm R}, 3/8},\\ \nonumber
f_3 &= \chi^{\mathrm{SM}_{\rm eff}(28,2) }_{{\rm NS},1/14}- \chi^{\mathrm{SM}_{\rm eff}(28,2) }_{{\rm NS},15/14}+2\chi^{\mathrm{SM}_{\rm eff}(28,2) }_{{\rm R}, 53/56}.
\end{align}
The Nahm sum associated with $(T_1,A_3)$ is identical to $f_1$ and thus can be expressed as
\begin{align}\label{eq:T1A3}
f_{A,\vec{0},-\frac{3}{56}}=\chi^{\mathrm{SM}_{\rm eff}(28,2) }_{{\rm NS},0}+\chi^{\mathrm{SM}_{\rm eff}(28,2) }_{{\rm NS},3/2}+2\chi^{\mathrm{SM}_{\rm eff}(28,2) }_{{\rm R}, 3/8}.
\end{align}

\vspace{3mm}

\textbf{Case $(A_1, A_2)$}.
This is Example 10 in Zagier's list \cite{Zag07} of rank two. In this case, $A=\begin{psmallmatrix}
4/3 & 2/3 \\ 2/3 &  4/3    
\end{psmallmatrix}$ and $c=4/5$. There are three Nahm sums, and we label them by $f_j$ for $1\leq j\leq 3$ in the order of Zagier's list. Note that $f_1=f_2$ and $f_3$ is the Nahm sum with $B=0$. We find that this case is related to $\mathrm{M}_{\mathrm{sub}}(6,5)$ by
\begin{align}\label{eq:A1A21}
f_1 &= 2\chi_{1/15}^{\mathrm{M}_{\mathrm{sub}}(6,5)} + \chi_{2/5}^{\mathrm{M}_{\mathrm{sub}}(6,5)},\\ \label{eq:A1A22}
f_3 &= \chi_0^{\mathrm{M}_{\mathrm{sub}}(6,5)} + 2\chi_{2/3}^{\mathrm{M}_{\mathrm{sub}}(6,5)}. 
\end{align}
These characters can be found, for example, in \cite[Section 5.5]{Duan:2022ltz}. 
The above coefficients 2 are consistent with the degeneracy of $\mathrm{M}_{\mathrm{sub}}(6,5)$.

\vspace{3mm}

\textbf{Case $(A_2, A_1)$}.
This is Example 11 in Zagier's list \cite{Zag07} of rank two. In this case, $A=\begin{psmallmatrix}
1 & -1/2 \\ -1/2 & 1    
\end{psmallmatrix}$ and $c=6/5$. We find that the associated CFT is $D_{\rm 2A}$, which is related to the 2A conjugacy class of the Monster group. There are three Nahm sums, and we label them by $f_j$ for $1\leq j\leq 3$ in the order of Zagier's list. Note that $f_1=f_2$ and $f_3$ is the Nahm sum with $B=0$.  We find the following relations
\begin{align}\label{eq:A2A11}
f_1 &= 3\chi_{1/10}^{D_{\rm 2A}} + \chi_{3/5}^{D_{\rm 2A}},\\ \label{eq:A2A12}
f_3 &= \chi_0^{D_{\rm 2A}} + 3\chi_{1/2}^{D_{\rm 2A}}. 
\end{align}
The above coefficients 3 are consistent with the degeneracy of $D_{\rm 2A}$ CFT. This theory is discussed in, e.g., \cite[Section 5.3]{Duan:2022ltz} denoted as $M_{6/5}$. It can be viewed as a non-diagonal modular invariant 
of $\text{Ising}\otimes \mathrm{M}(5,4)$.

\vspace{3mm}

\textbf{Case $(A_1, A_3)$}. 
As mentioned in Case (f) of Section \ref{sec:known-cases}, this case reduces to Example 1 in Zagier's list \cite{Zag07} of rank two with $\alpha=3/2$. The associated CFT is $U(1)_3$. We find the following relation between the Nahm sum and the $U(1)_3$ characters:
\begin{align}\label{eq:A1A3}
f_{A,\vec{0},-\frac{1}{24}}=\chi_0^{U(1)_3} +\chi_{3/4}^{U(1)_3} .
\end{align}
The Nahm sum can also be expressed by the characters of $\mathbb{Z}_4$ parafermion CFT \cite{KN11} as 
\begin{align}\label{eq:A1A32}
f_{A,\vec{0},-\frac{1}{24}}=\chi_0^{\mathbb{Z}_4} +\chi_1^{\mathbb{Z}_4} +2\chi_{3/4}^{\mathbb{Z}_4}. 
\end{align}

\vspace{3mm}

\textbf{Case $(A_1, B_3)$}. This case has appeared in Mizuno's Table 3 \cite{Miz23} with
$$
A=  \left(
\begin{array}{ccc}
 2 & 2 & 2 \\
 2 & 4 & 4 \\
 1 & 2 & 3 \\
\end{array}
\right).
$$
Mizuno conjectured 10 modular Nahm sums for this $A$-matrix, which have been all proved by \cite{WW24b}. 
We find that the associated CFT is the unitary $N=1$ minimal model $\mathrm{SM}(8,6)$ with central charge $5/4$. This supersymmetric CFT has nine NS weights
$
0,\frac{1}{32},\frac{1}{12},\frac{5}{32},\frac{1}{4},\frac{5}{6},\frac{33}{32},\frac{5}{4},3
$ and nine R weights 
$
\frac{5}{96},\frac{1}{16},\frac{3}{32},\frac{5}{16},\frac{41}{96},\frac{9}{16},\frac{23}{32},\frac{29}{16},\frac{67}{32}
$. 
We find the following relations between the Nahm sums and CFT characters:
\begin{align}\label{eq:A1B3}
   f_{A,(0,0,0),-\frac{5}{96}}&=\chi^{\mathrm{SM}(8,6)}_{\mathrm{NS},0}-\chi^{\mathrm{SM}(8,6)}_{\mathrm{NS},3},\\ \label{eq:A1B32}
   f_{A,(0,-1,-1),-\frac{1}{48}}&=\chi^{\mathrm{SM}(8,6)}_{\mathrm{NS},\frac{1}{32}}-\chi^{\mathrm{SM}(8,6)}_{\mathrm{NS},\frac{33}{32}},\\ \label{eq:A1B33}
   f_{A,(-1,0,1/2),\frac{1}{24}}&=\chi^{\mathrm{SM}(8,6)}_{\mathrm{R},\frac{3}{32}}-\chi^{\mathrm{SM}(8,6)}_{\mathrm{R},\frac{67}{32}}.
\end{align}

Now we present some evidence for Conjecture \ref{conj:A1}. The unfolded Nahm sum is of type $(A_1,A_5)$ and the associated CFT is the $\mathbb{Z}_6$ parafermion with central charge $5/4$.  
For $(A_1,A_5)$, the Nahm sum can be expressed by the characters of the $\mathbb{Z}_6$ parafermion CFT as 
\begin{align}\label{eq:A1A5}
f_{2C_{A_5}^{-1},\vec{0},-\frac{5}{96}}=\chi_0^{\mathbb{Z}_6} +\chi_{\frac32}^{\mathbb{Z}_6} +2\chi_{\frac56}^{\mathbb{Z}_6}+2\chi_{\frac43}^{\mathbb{Z}_6}. 
\end{align}
On the other hand, this Nahm sum can also be expressed by the characters of $\mathrm{SM}(8,6)$ as 
\begin{align}\label{eq:A1A52}
f_{2C_{A_5}^{-1},\vec{0},-\frac{5}{96}}=\chi^{\mathrm{SM}(8,6)}_{\mathrm{NS},0}+\chi^{\mathrm{SM}(8,6)}_{\mathrm{NS},3}+2\chi^{\mathrm{SM}(8,6)}_{\mathrm{NS},\frac56}. 
\end{align}
Given that the characters of the $\mathbb{Z}_6$ parafermion are $q$-series, while the NS characters of $\mathrm{SM}(8,6)$ are $\sqrt{q}$-series, it is easy to see that all the characters of the $\mathbb{Z}_6$ parafermion in \eqref{eq:A1A5} can be expressed as linear combinations of the characters of the bosonization of $\mathrm{SM}(8,6)$. We also checked that this is true for other characters of the $\mathbb{Z}_6$ parafermion. Therefore, the $\mathbb{Z}_6$ parafermion is indeed a non-diagonal modular invariant  of the bosonization of $\mathrm{SM}(8,6)$, which seems to be unknown in the literature. This is a rather nontrivial support for Conjecture \ref{conj:A1}.

\vspace{3mm}

\section*{Appendix: Dynkin diagrams and Cartan matrices}
For convenience we collect data from \cite{Bou68} for Dynkin diagrams, including the Cartan matrices, the Coxeter numbers, and the diagonal matrices $D(X)$ introduced in Conjecture \ref{Conj:main}. 

\begin{enumerate}
\item The Dynkin diagram $A_r$ for $r\geq 1$ has the Coxeter number $h(A_r)=r+1$, $D(A_r)=\mathrm{Id}_r$, and the Cartan matrix 
$$
C(A_r)=\begin{pmatrix}
2 & -1 & 0  & \cdots & 0 & 0 \\
-1 & 2 & -1 & \cdots & 0 & 0 \\
0 & -1 &  2 & \cdots & 0 & 0 \\
\vdots & \vdots & \vdots & \ddots & \vdots & \vdots  \\
0 & 0 & 0 & \cdots & 2 & -1 \\
0 & 0 & 0 & \cdots & -1 & 2
\end{pmatrix}.
$$

\item The Dynkin diagram $B_r$ for $r\geq 3$ has the Coxeter number $h(B_r)=2r$, 
$$
D(B_r)=\mathrm{diag}(2,...,2,1),
$$
and the Cartan matrix 
$$
C(B_r)=\begin{pmatrix}
2 & -1 & 0  & \cdots & 0 & 0 \\
-1 & 2 & -1 & \cdots & 0 & 0 \\
0 & -1 &  2 & \cdots & 0 & 0 \\
\vdots & \vdots & \vdots & \ddots & \vdots & \vdots  \\
0 & 0 & 0 & \cdots & 2 & -2 \\
0 & 0 & 0 & \cdots & -1 & 2
\end{pmatrix}.
$$

\item The Dynkin diagram $C_r$ for $r\geq 2$ has the Coxeter number $h(C_r)=2r$, 
$$
D(C_r)=\mathrm{diag}(1,...,1,2),
$$
and the Cartan matrix 
$$
C(C_r)=\begin{pmatrix}
2 & -1 & 0  & \cdots & 0 & 0 \\
-1 & 2 & -1 & \cdots & 0 & 0 \\
0 & -1 &  2 & \cdots & 0 & 0 \\
\vdots & \vdots & \vdots & \ddots & \vdots & \vdots  \\
0 & 0 & 0 & \cdots & 2 & -1 \\
0 & 0 & 0 & \cdots & -2 & 2
\end{pmatrix}.
$$

\item The Dynkin diagram $D_r$ for $r\geq 4$ has the Coxeter number $h(D_r)=2r-2$, $D(D_r)=\mathrm{Id}_r$, and the Cartan matrix 
$$
C(D_r)=\begin{pmatrix}
2 & -1 & 0  & \cdots & 0 & 0 & 0 \\
-1 & 2 & -1 & \cdots & 0 & 0 & 0 \\
0 & -1 &  2 & \cdots & 0 & 0 & 0 \\
\vdots & \vdots & \vdots & \ddots & \vdots & \vdots & \vdots  \\
0 & 0 & 0 & \cdots & 2 & -1 & -1  \\
0 & 0 & 0 & \cdots & -1 & 2 & 0 \\
0 & 0 & 0 & \cdots & -1 & 0 & 2
\end{pmatrix}.
$$

\item The Dynkin diagram $E_6$ has the Coxeter number $h(E_6)=12$, $D(E_6)=\mathrm{Id}_6$, and the Cartan matrix 
$$
C(E_6)=\begin{pmatrix}
2 & 0 & -1  & 0 & 0 & 0  \\
0 & 2 & 0 & -1 & 0 & 0  \\
-1 & 0 & 2 & -1 & 0 & 0  \\
0 & -1 & -1 & 2 & -1 & 0  \\
0 & 0 & 0 & -1 & 2 & -1  \\
0 & 0 & 0 & 0 & -1 & 2  
\end{pmatrix}.
$$

\item The Dynkin diagram $E_7$ has the Coxeter number $h(E_7)=18$, $D(E_7)=\mathrm{Id}_7$, and the Cartan matrix 
$$
C(E_7)=\begin{pmatrix}
2 & 0 & -1  & 0 & 0 & 0 & 0 \\
0 & 2 & 0 & -1 & 0 & 0 & 0 \\
-1 & 0 & 2 & -1 & 0 & 0 & 0 \\
0 & -1 & -1 & 2 & -1 & 0 & 0 \\
0 & 0 & 0 & -1 & 2 & -1 & 0 \\
0 & 0 & 0 & 0 & -1 & 2 & -1 \\
0 & 0 & 0 & 0 & 0 & -1 & 2
\end{pmatrix}.
$$

\item The Dynkin diagram $E_8$ has the Coxeter number $h(E_8)=30$, $D(E_8)=\mathrm{Id}_8$, and the Cartan matrix 
$$
C(E_8)=\begin{pmatrix}
2 & 0 & -1  & 0 & 0 & 0 & 0 & 0 \\
0 & 2 & 0 & -1 & 0 & 0 & 0 & 0 \\
-1 & 0 & 2 & -1 & 0 & 0 & 0 & 0 \\
0 & -1 & -1 & 2 & -1 & 0 & 0 & 0\\
0 & 0 & 0 & -1 & 2 & -1 & 0 & 0 \\
0 & 0 & 0 & 0 & -1 & 2 & -1 & 0 \\
0 & 0 & 0 & 0 & 0 & -1 & 2 & -1 \\
0 & 0 & 0 & 0 & 0 & 0 & -1 & 2
\end{pmatrix}.
$$

\item The Dynkin diagram $F_4$ has the Coxeter number $h(F_4)=12$, 
$$
D(F_4)=\mathrm{diag}(2,2,1,1),
$$
and the Cartan matrix 
$$
C(F_4)=\begin{pmatrix}
2 & -1 & 0 & 0\\
-1 & 2 & -2 & 0 \\
0 & -1 & 2 & -1\\
0 & 0 & -1 & 2
\end{pmatrix}.
$$

\item The Dynkin diagram $G_2$ has the Coxeter number $h(G_2)=6$, 
$$
D(G_2)=\mathrm{diag}(1,3),
$$
and the Cartan matrix 
$$
C(G_2)=\begin{pmatrix}
2 & -1 \\
-3 & 2 
\end{pmatrix}.
$$

\item The Dynkin diagram $T_r$ for $r\geq 1$ has the Coxeter number $h(T_r)=2r+1$, $D(T_r)=\mathrm{Id}_r$, and the Cartan matrix 
$$
C(T_r)=\begin{pmatrix}
2 & -1 & 0  & \cdots & 0 & 0 \\
-1 & 2 & -1 & \cdots & 0 & 0 \\
0 & -1 &  2 & \cdots & 0 & 0 \\
\vdots & \vdots & \vdots & \ddots & \vdots & \vdots  \\
0 & 0 & 0 & \cdots & 2 & -1 \\
0 & 0 & 0 & \cdots & -1 & 1
\end{pmatrix}.
$$
\end{enumerate}

\bibliographystyle{plainnat}
\bibliofont

\end{document}